\documentclass[aps,preprintnumbers,superscriptaddress,showpacs]{revtex4}
\usepackage{epsfig}
\usepackage{psfrag}
\usepackage{amsfonts}
\usepackage{graphicx}
\usepackage{dcolumn}
\usepackage{bm}

\begin{document}

\title{Entropic destruction of heavy quarkonium in heavy quark cloud}

\author{Zi-qiang Zhang}
\email{zhangzq@cug.edu.cn} \affiliation{School of Mathematics and
Physics, China University of Geosciences, Wuhan 430074, China}

\begin{abstract}
Previous research has shown that the peak of the quarkonium
entropy at the deconfinement transition would be related to the
entropic force which induces the melting of quarkonium. In this
article, we study the effect of backreaction on the entropic force
in a strongly coupled plasma of adjoint matter. The backreaction
covered here comes from the presence of static heavy quarks evenly
distributed over such a plasma. It is found that the inclusion of
backreaction increases the entropic force thus enhancing the
quarkonium dissociation, in accord with the findings of the
imaginary potential.

\end{abstract}
\pacs{11.25.Tq, 11.15.Tk, 11.25-w} \maketitle
\section{Introduction}
It is believed that the high energy heavy-ion collisions at
Relativistic Heavy Ion Collider (RHIC) and Large Hadron Collider
(LHC) have created a strongly coupled hot quark-gluon plasma (QGP)
\cite{JA,KA,EV}. One of the main experimental signatures for QGP
formation is quarkonium dissociation \cite{TMA}. It is expected to
be produced during the early stages of the collisions and gives us
important information about the entire evolution of QGP, e.g., it
was predicted that the quarkonium would be suppressed due to the
Debye screening induced by the high density of color charges in
QGP. But recently, the experimental study on charmonium
($c\bar{c}$) shows a puzzle: the $c\bar{c}$ suppression at the
RHIC (lower energy density) seems to be stronger than that at LHC
(larger energy density) \cite{AAD,BBA}. Obviously, this is in
contradiction with the Debye screening as well as the thermal
activation through the impact of gluons \cite{DKH,EV1}. To explain
this, some scholars argued \cite{PBR,RLT} that the recombination
of the produced charm quarks into charmonium may be a solution. In
particular, if a region of deconfined quarks and gluons is formed,
the charmonium could be formed from a charm and an anticharm which
were originally produced in separate incoherent interactions.

However, D. Kharzeev has recently argued \cite{DEK} that the
puzzle on the suppression of $c\bar{c}$ may be related to the
nature of deconfinement. Specifically, the peak of the quarkonium
entropy at the deconfinement transition could be related to the
entropic force which induces the dissociation of quarkonium. This
argument is based on the Lattice results \cite{DKA1,DKA2,PPE,OK1}
showing a large amount of entropy $S$ associated with the heavy
quark-antiquark pair ($q\bar{q}$) around the crossover region of
QGP. In the proposal of \cite{DEK}, this entropy gives rise to the
entropic force
\begin{equation}
\mathcal{F}=T\frac{\partial S}{\partial L},\label{f}
\end{equation}
where $T$ is the temperature of the plasma and $L$ denotes the
interquark distance of $q\bar{q}$. It should be noted that this
force does not describe any additional interactions; instead, it
is an emergent force that originates from multiple interactions,
driving the system towards the state with a larger entropy. The
entropic force was introduced for the first time in \cite{KHM} to
explain the elasticity of polymer strands in rubber and recently
argued \cite{EPV} to be responsible for gravity. Here we will not
go into detail about these points and restrict our discussion to
its application in quarkonium dissociation.

AdS/CFT \cite{Maldacena:1997re,Gubser:1998bc,EW}, namely the
duality between the type IIB superstring theory formulated on
$AdS_5\times S^5$ and $\mathcal N=4$ supersymmetric Yang-Mills
(SYM) in four dimensions, provides a new method for studying
different aspects of QGP (see \cite{JCA} for a good review with
many phenomenological applications). Using AdS/CFT, K. Hashimoto
and D. Kharzeev first studied the entropic force in $\mathcal N=4$
SYM plasma at finite temperature \cite{KHA}. Since then, this
quantity has been discussed in various holographic models, e.g.,
for moving quarkonium \cite{KBF}, with chemical potential
\cite{ZQ} and higher derivative corrections \cite{ZQ1}. Other
related results can be found in \cite{II,DE1,ZQ2,DD}.

The goal of this paper is to study the effect of backreaction on
the entropic force. As is well known, the QGP produced in
experiments is comprised of a large amount of free quarks and
gluons. That means if one analyzes the dynamics of a quark or
quarkonium, the effect of other heavy quarks due to the
backreaction of the plasma may have to be considered. However,
with rare exceptions, e.g., \cite{FBI,SP}, it remains hard to
treat strongly coupled gauge theory with a large number of flavor
quarks. Because of this possibility, in most previous studies of
hydrodynamical aspects associated with QGP, the backreaction of
the plasma is usually neglected. Recently, S. Chakrabortty
proposed \cite{sch} a backreacted gravity background, which is
parametrized by the mass of the black hole and long string
density. In particular, the backreacted geometry is realized as an
AdS black hole back reacted in the inclusion of a uniform
distribution of large number of fundamental strings. It turns out
that this geometry is thermodynamically stable under tensor and
vector perturbations (see \cite{MHE} for a similar study on the
stability of the gravity configurations from the free energy
calculation). Subsequently, the drag force \cite{sch} and jet
quenching parameter \cite{sch1} have been studied in such a model.
It is shown that the presence of backreaction increases the drag
force and jet quenching parameter thus enhancing the energy loss.
More recently, the imaginary potential \cite{ZQ3} of a heavy
quarkonium was considered in the same model and the results show
that the inclusion of backreaction decreases the thermal width
thus enhancing the quarkonium dissociation. Inspired by these
facts, we wonder how does backreaction affect the entropic force?
Does backreaction have the same effect on the quarkonium
dissociation associated with the entropic force as with the
imaginary potential? We are answering these questions in the
present work.

The structure of the paper is as follows. In the next section, we
briefly review the backreacted gravity geometry given in
\cite{sch}. In section 3, we investigate the behavior of the
entropic force for this background and discuss how backreaction
influences the quarkonium dissociation. Finally, we conclude with
a discussion in section 4.

\section{background geometry}
One considers the (n+1)-dimensional gravitational action as
follows \cite{sch}
\begin{equation}
I=\frac{1}{16\pi
G_{n+1}}\int{dx^5\sqrt{-g}(\mathcal{R}-2\Lambda)+S_m}, \label{l}
\end{equation}
where $G_{n+1}$ is the (n+1)-dimensional Newton constant.
$\mathcal{R}$ denotes the Ricci scalar. $\Lambda$ represents the
negative cosmological constant. $S_m$ refers to the matter part,
\begin{equation}
S_m=-\frac{1}{2}\sum_i{\mathcal{T}_i}\int
d^2\xi\sqrt{-h}h^{\alpha\beta}\partial_\alpha X^\mu \partial_\beta
X^\nu g_{\mu\nu},
\end{equation}
where $h_{\alpha\beta}$ is the world-sheet metric while
$g_{\mu\nu}$ denotes the space-time metric, with $\alpha$, $\beta$
the world-sheet coordinates and $\mu$, $\nu$ the space-time
directions.

The Einstein's equations obtained from (\ref{l}) are
\begin{equation}
R_{\mu\nu}-\frac{1}{2}Rg_{\mu\nu}+\Lambda g_{\mu\nu} =8\pi G_{n+1}
T_{\mu\nu},
\end{equation}
with
\begin{equation}
T^{\mu\nu}=-\sum_i\mathcal{T}_i\int
d^2\xi\frac{1}{\sqrt{|g_{\mu\nu}|}}\sqrt{|h_{\alpha\beta}|}h^{\alpha\beta}\partial_\alpha
X^\mu\partial_\beta X^\nu{\delta_i}^{n+1}(x-X),
\end{equation}
where the delta function represents the source divergences due to
the inclusion of the strings.

Following \cite{sch}, the ansatz for the geometry can be written
as
\begin{equation}
ds^2=g_{tt}(r)dt^2+g_{rr}(r)dr^2+r^2\delta_{ab}dx^adx^b,
\end{equation}
where (a,b) run over n-1 space directions.

Choosing the static gauge $t=\xi^0$, $r=\xi^1$, then the
nonvanishing components of $T^{\mu\nu}$ become
\begin{equation}
T^{tt}=-\frac{ag^{tt}}{r^{n-1}},\qquad
T^{rr}=-\frac{ag^{rr}}{r^{n-1}},
\end{equation}
where the strings are supposed to be uniformly distributed over
n-1 directions, such that the string cloud density reads
\begin{equation}
a(x)=T\sum_i\delta_i^{(n-1)}(x-X_i), \qquad with \qquad a>0.
\end{equation}

Solving Einstein's equations, one gets
\begin{equation}
V(r)=K+\frac{r^2}{R^2}-\frac{2m}{r^{n-1}}-\frac{2a}{(n-1)r^{n-3}},
\end{equation}
where $R$ denotes the AdS radius. $K=0,-1,1$ correspond to the
boundary being flat, spherical or hyperbolic, respectively.

In this work, we are mostly interested in the case of $K=0, n=4$.
Given that, the metric becomes
\begin{equation}
ds^2=\frac{r^2}{R^2}(-f(r)dt^2+d\vec{x}^2)+\frac{R^2}{r^2f(r)}dr^2,\label{S0}
\end{equation}
with
\begin{equation}
f(r)=1-\frac{2mR^2}{r^4}-\frac{2}{3}\frac{aR^2}{r^3},
\end{equation}
where $r$ is the $5th$ dimensional coordinate with $r=\infty$ the
boundary and $r=r_h$ the horizon, where $r_h$ satisfies
$f(r_h)=0$.

The parameter $m$ reads
\begin{equation}
m=\Big(1-\frac{2}{3}\frac{aR^2}{{r_h}^3}\Big)\frac{{r_h}^4}{2R^2}.
\end{equation}

The temperature of the black hole reads
\begin{equation}
T=\frac{\sqrt{g^{rr}}\partial_r\sqrt{g_{tt}}}{2\pi}|_{r=r_h}=\frac{6{r_h}^3-aR^2}{6\pi
R^2{r_h}^2}.\label{T0}
\end{equation}

As shown in \cite{sch}, the geometry (\ref{S0}) is
thermodynamically stable under tensor and vector perturbations.
Moreover, it resembles an AdS Schwarzschild black hole with
negative curvature horizon. For more information about it, refer
to \cite{sch}.

\section{entropic force in the backreacted gravity background}
In this section we follow the prescription in \cite{KHA} to study
the behavior of the entropic force for the background metric
(\ref{S0}).

The Nambu-Goto action is
\begin{equation}
S_{NG}=-\frac{1}{2\pi\alpha^\prime}\int d\tau d\sigma\mathcal
L=-\frac{1}{2\pi\alpha^\prime}\int d\tau
d\sigma\sqrt{-detg_{\alpha\beta}}, \label{S}
\end{equation}
with
\begin{equation}
g_{\alpha\beta}=g_{\mu\nu}\frac{\partial
X^\mu}{\partial\sigma^\alpha} \frac{\partial
X^\nu}{\partial\sigma^\beta},
\end{equation}
where $g_{\alpha\beta}$ is the induced metric, parametrized by
$(\tau,\sigma)$ on the string world-sheet. $X^\mu$ denotes the
target space coordinate.

Taking the static gauge
\begin{equation}
t=\tau, \qquad x_1=\sigma,
\end{equation}
and supposing $r$ depends only on $\sigma$,
\begin{equation}
r=r(\sigma),
\end{equation}
then the Lagrangian density becomes
\begin{equation}
\mathcal L=\sqrt{\frac{r^4f(r)}{R^4}+\dot{r}^2},\label{L}
\end{equation}
with $\dot{r}\equiv dr/d\sigma$.

Since $\mathcal L$ does not depend on $\sigma$ explicitly, one has
a conserved quantity,
\begin{equation}
\mathcal L-\frac{\partial\mathcal
L}{\partial\dot{r}}\dot{r}=constant.
\end{equation}

Imposing the boundary condition at $\sigma=0$,
\begin{equation}
\dot{r}=0,\qquad  r=r_c \qquad (r_h<r_c)\label{con},
\end{equation}
the conserved quantity turns into
\begin{equation}
\frac{r^4f(r)/R^4}{\sqrt{\frac{r^4f(r)}{R^4}+\dot{r}^2}}=\sqrt{r_c^4f(r_c)/R^4},
\end{equation}
yielding
\begin{equation}
\dot{r}=\frac{dr}{d\sigma}=\sqrt{\frac{A^2(r)-A(r)A(r_c)}{A(r_c)}}\label{dotr},
\end{equation}
with
\begin{equation}
A(r)=\frac{r^4f(r)}{R^4},\qquad A(r_c)=\frac{r_c^4f(r_c)}{R^4},
\qquad f(r_c)=1-\frac{2mR^2}{r_c^4}-\frac{2}{3}\frac{aR^2}{r_c^3}.
\end{equation}

Integrating (\ref{dotr}), the interdistance of $q\bar{q}$ is
obtained as
\begin{equation}
L=2\int_{r_c}^{\infty}\frac{d\sigma}{dr}dr=2\int_{r_c}^{\infty}dr\sqrt{\frac{A(r_c)}{A^2(r)-A(r)A(r_c)}}\label{x}.
\end{equation}

In fig. 1, we plot $LT$ as a function of $\varepsilon$ for
different values of $a$, where $\varepsilon\equiv r_h/r_c$ and
$0<\varepsilon<1$. From these figures, one can see that increasing
$a$ leads to decreasing $LT$, indicating the quark and antiquark
get closer under the influence of backreaction.

\begin{figure}
\centering
\includegraphics[width=10cm]{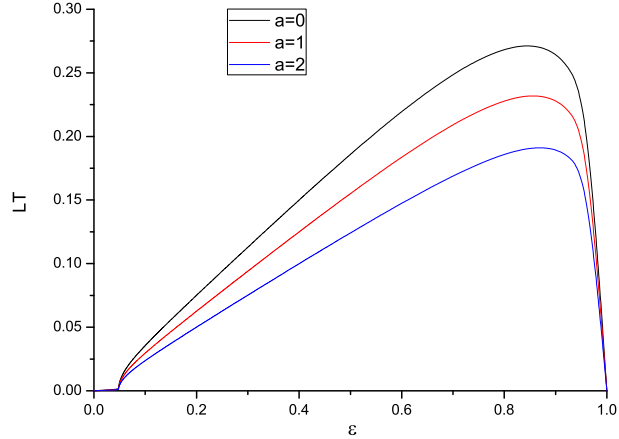}
\caption{$LT$ versus $\varepsilon$ for different values of $a$.
From top to bottom, $a=0,1,2$, respectively.}
\end{figure}

The next step is to calculate the entropy $S$, which takes the
form
\begin{equation}
S=-\frac{\partial F}{\partial T},\label{s}
\end{equation}
where $F$ represents the free energy of $q\bar{q}$ (see
\cite{JMM,ABR,SJR} for the original calculation of the rectangular
Wilson loop associated with free energy in the vacuum of strongly
coupled $\mathcal N=4$ SYM and its generalization to finite
temperature). In general, there are two situations:

1. If $L>\frac{c}{T}$ (with $c$ the maximum value of $LT$), the
quarks are completely screened. For this case, one needs to
consider some new configurations \cite{DB} such that the choice of
the $F$ is not unique \cite{MCH}. Here we choose a configuration
of two disconnected trailing drag strings \cite{CPH,GB}. At this
point, the free energy is
\begin{equation}
F^{(1)}=\frac{1}{\pi\alpha^\prime}\int_{r_h}^{\infty}dr,
\end{equation}
yielding
\begin{equation}
S^{(1)}=\sqrt{\lambda}\theta(L-\frac{c}{T})\label{S1},
\end{equation}
where $\theta(L-\frac{c}{T})$ refers to the Heaviside step
function.

2. If $x<\frac{c}{T}$, the fundamental string is connected. For
this case, the free energy equals the on-shell action of the
fundamental string in the dual geometry,
\begin{equation}
F^{(2)}=\frac{1}{\pi\alpha^\prime}\int_{r_c}^{\infty} dr
\sqrt{\frac{A(r)}{A(r)-A(r_c)}}.\label{S2}
\end{equation}

Since $m$ is related to $r_h$, one may rewrite Eq.(\ref{s}) as
\begin{equation}
S=-\frac{\partial F}{\partial T}=-\frac{\partial F}{\partial
r_h}/\frac{\partial T}{\partial r_h}=-(\frac{1}{\pi
}+\frac{a}{3\pi r_h^3})\frac{\partial F}{\partial r_h},\label{s1}
\end{equation}
where, for convenience, we have set $R=1$.

As a result, one gets
\begin{equation}
S^{(2)}=-\frac{1}{\pi\alpha^\prime}(\frac{1}{\pi}+\frac{a}{3\pi
r_h^3})\int_{r_c}^{\infty}
dr\frac{A^\prime(r)[A(r)-A(r_c)]-A(r)[A^\prime(r)-A^\prime(r_c)]}{\sqrt{A(r)[A(r)-A(r_c)]^3}}\label{S21},\label{fo}
\end{equation}
with
\begin{eqnarray}
A^\prime(r)&=&r^4f^\prime(r), \qquad f^\prime(r)=-2r^{-4}(2r_h^3-\frac{a}{3}), \nonumber\\
A^\prime(r_c)&=&r_c^4f^\prime(r_c), \qquad
f^\prime(r_c)=-2r_c^{-4}(2r_h^3-\frac{a}{3}),
\end{eqnarray}
where the derivatives are with respect to $r_h$. Note that by
plugging $a=0$ in Eq.(\ref{fo}), the result of SYM \cite{KHA} is
reproduced.

Now we analyze how backreaction modifies the entropic force. To
that end, we plot $S^{(2)}/\sqrt{\lambda}$ as a function of $LT$
for different values of $a$ in fig.2, where we have used the
relation $\alpha^\prime=1/\sqrt{\lambda}$. From these figures, it
is seen that increasing $a$ leads to larger entropy at small
distances. As you know, the entropic force is found to grow as a
function of the distance (see Eq.(\ref{f})) and responsible for
melting the quarkonium. Consequently, one concludes that the
inclusion of backreaction increases the entropic force thus
enhancing the quarkonium dissociation, in agreement with the
findings of the imaginary potential \cite{ZQ3}.

However, it should be pointed out that the temperature depends on
$r_h$ and $a$ (see Eq.(\ref{T0})), thus, fig.2 actually involves
the contributions of temperature variation as well. However, the
existing research, e.g, \cite{ZQ2}, shows that as $T$ increases
the entropic force increases. On that basis one reasons as
follows: the overall result is that increasing $a$ leads to
increasing the entropic force, but at the same time increasing $a$
leads to decreasing $T$ thus decreasing the entropic force.
Therefore, one infers that with fixed $T$, increasing $a$
necessarily increases the entropic force. The physical meaning of
the results will be discussed in the next section.

\begin{figure}
\centering
\includegraphics[width=10cm]{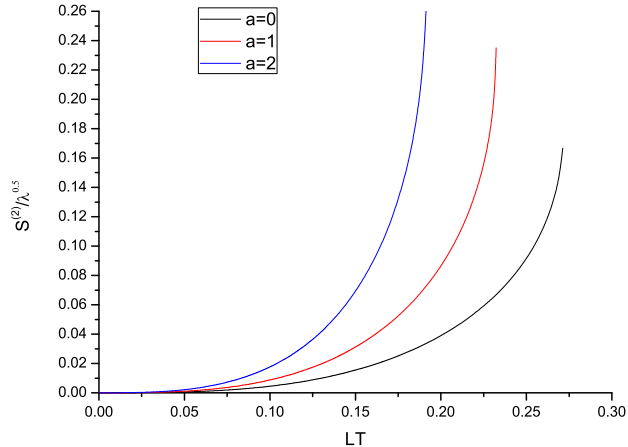}
\caption{$S^{(2)}/\sqrt{\lambda}$ versus $LT$ for different values
of $a$. From top to bottom $a=2,1,0$, respectively.}
\end{figure}

\section{conclusion}
Recent research has suggested that the entropic force may
represent a mechanism for dissociating the heavy quarkonium. In
this paper, we studied the effect of backreaction on the entropic
force in a strongly coupled plasma of adjoint matter. This
backreaction is thought to come from the presence of static heavy
quarks evenly distributed over such a plasma. It turns out that
the inclusion of backreaction increases the entropic force thus
enhancing the quarkonium dissociation, consistently with the
findings of the imaginary potential.

However, there are many places worth further improvement and
in-depth study. For example, how to analyze the backreaction
effect associated with the interaction between the heavy quarks in
the cloud and the backreaction to the spacetime geometry? (Of
course this case is more complicated.) Also, how to study such
effect with respect to nonuniform distributed quarks? (It is quite
possible that the quarks won't be evenly distributed in the
experiments.) Moreover, how to consider such effect in
nonconformal systems? We hope to tackle these issues in the near
future.

\section{Acknowledgments}
This work is supported by the NSFC under Grant No. 11705166 and
the Fundamental Research Funds for the Central Universities, China
University of Geosciences (Wuhan) (No. CUGL180402).

\end{document}